\documentclass[twocolumn, times, trackchanges]{aastex63_bren}
\usepackage{CJK}
\usepackage{booktabs} 
\usepackage{xcolor, soul}
\usepackage{amsmath, amssymb, bm}

\newcommand{\WDmodelsurl}{\href{https://github.com/SihaoCheng/WD_models}{\url{https://github.com/SihaoCheng/WD_models}}}

\newcommand{\DWDmergersurl}{\href{https://sihaocheng.github.io/DWDmergers}{\url{https://sihaocheng.github.io/DWDmerger}}}

\newcommand{\wdurl}{\href{http://www.astro.umontreal.ca/~bergeron/CoolingModels/}{\url{http://www.astro.umontreal.ca/~bergeron/CoolingModels/}}}

\newcommand{\vL}{v_l}
\newcommand{\vB}{v_b}
\newcommand{\mathbfit}{\bm}
\newcommand{\bvT}{\mathbfit{v}_\text{T}}

\begin{document}
\begin{CJK*}{UTF8}{gbsn}
\received{2019 November 12}
\revised{2020 February 2}
\accepted{2020 February 4}
\submitjournal{\apj}

\title{Double White Dwarf Merger Products among High-mass White Dwarfs}
\shorttitle{Merger Products among WDs}
\shortauthors{Cheng, Cummings, M\'enard, \& Toonen}

\email{s.cheng@jhu.edu}

\author[0000-0002-9156-7461]{Sihao Cheng (程思浩)}
\affiliation{Department of Physics and Astronomy, The Johns Hopkins University, 3400 N Charles Street, Baltimore, MD 21218, USA}

\author[0000-0001-7453-9947]{Jeffrey D. Cummings}
\affiliation{Department of Physics and Astronomy, The Johns Hopkins University, 3400 N Charles Street, Baltimore, MD 21218, USA}

\author{Brice M\'enard}
\affiliation{Department of Physics and Astronomy, The Johns Hopkins University, 3400 N Charles Street, Baltimore, MD 21218, USA}
\affiliation{Kavli Institute for the Physics and Mathematics of the Universe, University of Tokyo, Kashiwa 277-8583, Japan}

\author[0000-0002-2998-7940]{Silvia Toonen}
\affiliation{
Institute for Gravitational Wave Astronomy and School of Physics and Astronomy, University of Birmingham, Birmingham B15 2TT, UK}

\begin{abstract}
Double white dwarf (double-WD) binaries may merge within a Hubble time and produce high-mass WDs. Compared to other high-mass WDs, the double-WD merger products have higher velocity dispersion because they are older. With the power of {\it Gaia} data, we show strong evidence for double-WD merger products among high-mass WDs by analyzing the transverse-velocity distribution of more than a thousand high-mass WDs (0.8--1.3~$M_\odot$). We estimate that the fraction of double-WD merger products in our sample is about 20\%. We also obtain a precise double-WD merger rate and its mass dependence. Our merger rate estimates are close to binary population synthesis results and support the idea that double-WD mergers may contribute to a significant fraction of type Ia supernovae.
\end{abstract}

\keywords{White dwarf stars (1799); Stellar kinematics (1608); Stellar ages (1581); Type Ia supernovae (1728); Bayesian statistics (1900)}

\section{Introduction}
\label{sec:intro}

During the last few decades, there has been increasing evidence showing that a large number of double white dwarf (double-WD) systems should merge within a Hubble time \citep[e.g.,][]{Marsh_1995, Marsh_1995b, Iben_1996, Han_1998, Badenes_2012, Maoz_2018, Brown_2020}. Many double-WD mergers are believed to produce new white dwarfs with higher masses \citep[e.g.,][]{Loren-Aguilar_2009}. So, a fraction of high-mass white dwarfs in the solar neighborhood are expected to be double-WD merger products \citep[e.g.,][]{Toonen_2017, Temmink_2019}. To verify the existence of these merger products, some investigators have looked for an excess of high-mass white dwarfs \citep{Giammichele_2012, Rebassa-Mansergas_2015, Tremblay_2016}, and others have searched for kinematic signatures of these merger products \citep{Wegg_2012, Dunlap_2015}. The kinematic method makes use of the following facts: high-mass double-WD merger products are in general older than singly-evolved white dwarfs because of their binary evolution, and according to the age--velocity-dispersion relation (AVR) of the Milky-Way disc \citep[e.g.,][]{Nordstrom_2004}, these older double-WD merger products have higher velocity dispersion.
The former method, based on number counts, is influenced by large systematic errors from the adopted initial--final-mass relation of white dwarfs and the sample completeness. In contrast, the kinematic method is less influenced by systematic errors, but it was limited by the sample size of white dwarfs with kinematic measurements.

Thanks to the European Space Agency {\it Gaia} mission \citep{GaiaCollaboration_2016}, the number of stars with precise kinematic measurements has been enlarged drastically. \citet{Cheng_2019} selected a deep, homogeneous sample of white dwarfs in a narrow mass range (1.08--1.23~$M_\odot$) from {\it Gaia} Data Release 2 \citep[DR2;][]{GaiaCollaboration_2018a} to investigate the `Q branch', an overdensity of white dwarfs on the Hertzsprung--Russell (H--R) diagram, which is caused by a cooling anomaly. As a byproduct of their kinematic analysis, the fraction of double-WD merger products among white dwarfs in their mass range were inferred to be about 22\%, and they reserved the task of conducting an analysis optimized for detecting double-WD merger products and the discussion on this topic to this paper.

In this paper, we extend the kinematic analysis of high-mass white dwarfs to a wider mass range and adopt a more realistic delay-time distribution for binary evolution. We estimate the fractions of double-WD merger products as a function of mass and calculate the corresponding merger rates. We then compare our results to predictions from binary population synthesis. We also discuss the implication of our results for the progenitor problem of type Ia supernovae (SNe Ia), as the double-WD merger is a promising scenario of type Ia supernova explosions \citep[e.g.,][]{Iben_1984, Webbink_1984}.

\section{Data}
\label{sec:data}

{\it Gaia} DR2 provides accurate astrometric \citep{Lindegren_2018} and photometric \citep{Evans_2018, Riello_2018} measurements for more than one billion of stars. To search for the kinematic signature of double-WD merger products efficiently, we select nearby, high-mass, hot white dwarfs with precise astrometric and photometric measurements from the {\it Gaia} DR2 white dwarf catalog compiled by \citet{GentileFusillo_2019}. Below, we introduce in detail our sample selection and the derivation of white dwarf parameters. 

We first impose the same quality cuts as equations (1)--(5) in \citet{Cheng_2019} and a distance cut
\begin{align}
\label{eq:d_selection}
    1/\varpi<250\text{ pc}\, 
\end{align}
to select white dwarfs with high-precision astrometric and photometric measurements. These cuts do \emph{not} introduce any \emph{explicit} kinematic biases to our white dwarf sample. 

Then, as the kinematic signature of double-WD merger products most outstanding among high-mass, hot (young-photometric-age) white dwarfs, we carry out selections on the photometric mass ($m_\text{WD}$) and age ($\tau_\text{phot}$) of white dwarfs. These cuts, equivalent to cuts on the H--R diagram, are designed to both maximize the sample size and minimize contamination from the standard-mass helium-atmosphere white dwarfs \citep[the `B branch',][]{GaiaCollaboration_2018b}:
\begin{align}
\label{eq:m_t_selection}
    \nonumber 0.8\,M_\odot<&\,m_\text{WD}<0.9\,M_\odot\,,\\
    \nonumber 0.1\,\text{Gyr} < &\,\tau_\text{phot} < 0.7\,\text{Gyr}\,;\\
    \nonumber \text{or}\\
    \nonumber 0.9\,M_\odot<&\,m_\text{WD}<1.28\,M_\odot\,,\\
    0.1\,\text{Gyr} < &\,\tau_\text{phot} < 1.0\,\text{Gyr}\,.
\end{align}
The white dwarf parameters $m_\text{WD}$ and $\tau_\text{phot}$ are derived in the following way. First, we define the absolute magnitude $M_G$ as $M_G = G + 5\log(\varpi/\text{mas}^{-1}) - 10$, where $G$ and $\varpi$ are the $G$-band magnitude and parallax. Then, we convert the H--R diagram coordinate into $m_\text{WD}$ and cooling time $t_\text{cool}$ by interpolating a grid of cooling tracks for C/O-core DA (hydrogen atmosphere) white dwarfs \citep{Fontaine_2001} and synthetic colors \citep{Holberg_2006, Kowalski_2006, Tremblay_2011}.\footnote{\wdurl.}\footnote{We made a python 3 package for this kind of transformation publicly available on \WDmodelsurl.} For white dwarfs heavier than 1.07~$M_\odot$ we use cooling tracks of O/Ne white dwarfs \citep{Camisassa_2019}. Finally, the photometric age $\tau_\text{phot}$ is obtained by adding the cooling time $t_\text{cool}$ to the main-sequence age, which we calculate based on an initial--final mass relation \citep{Cummings_2018} and the relation between pre-cooling time and main-sequence mass from \citet{Choi_2016} for non-rotating, solar-metallicity stars.

\begin{figure}
    \centering
    \includegraphics[width=\columnwidth]{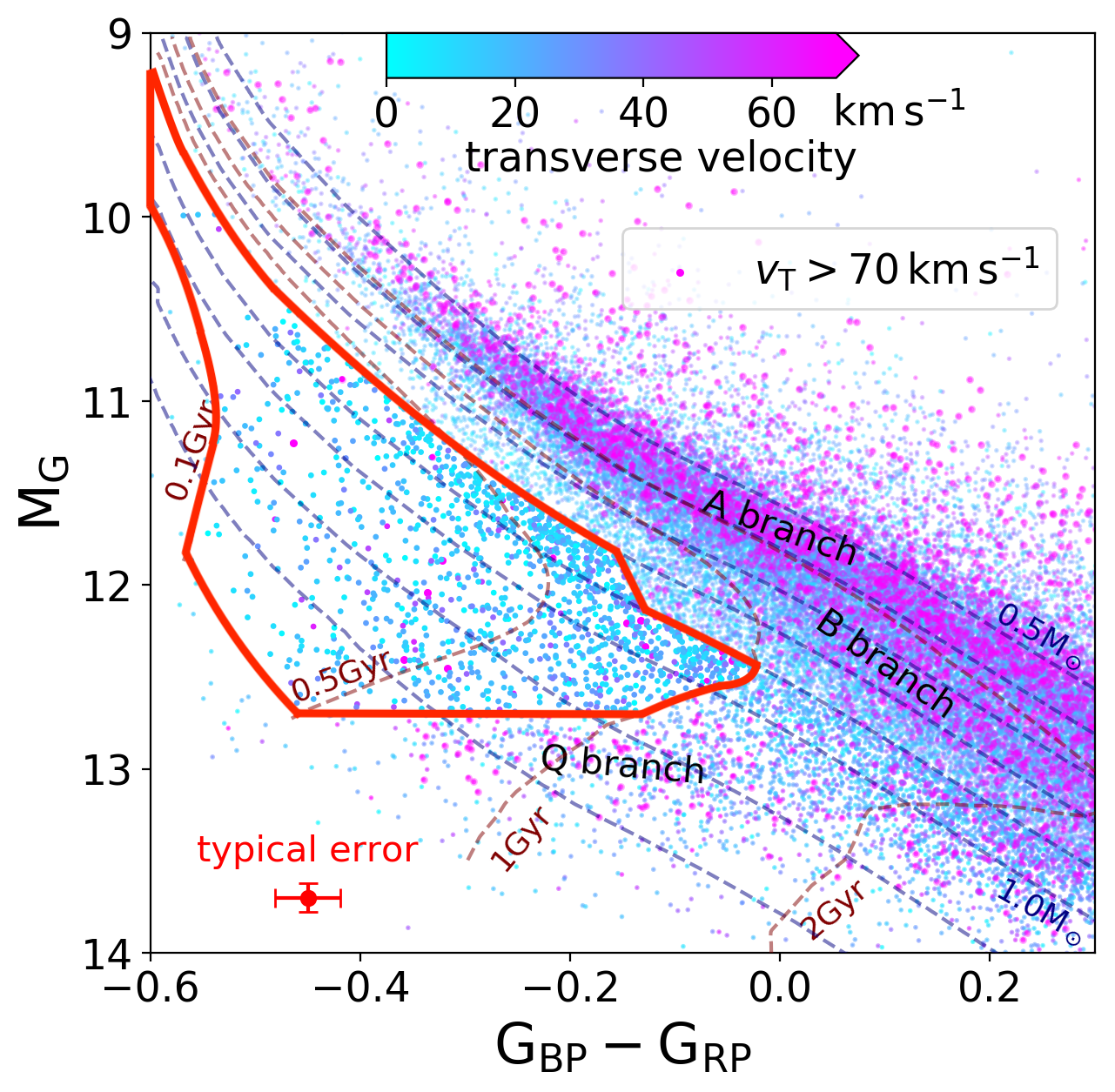}
    \caption{\label{fig:HR_merger} H--R diagram of WDs in {\it Gaia} DR2. We show the 250~pc sample of WDs with high-quality measurements and a grid of WD masses $m_\text{WD}$ and photometric ages $\tau_\text{phot}$ derived from the combined O/Ne- and C/O-core WD cooling model. WDs evolve along their cooling tracks, i.e., the constant-mass curves. The red region includes 1395 nearby, high-mass, hot WDs selected in Section~\ref{sec:data}.
    }
\end{figure}

As shown by \citet{Cheng_2019}, the `Q branch' on the H--R diagram is produced by an anomalous cooling behavior: some white dwarfs stop cooling and stay on the branch for several billion years, which creates both an overdensity and a high-velocity excess. To avoid modelling the influence of this cooling delay on the velocity distribution and only focus on the binary-evolution delay for double-WD mergers, we exclude the `Q branch' region on the H--R diagram by the cut
\begin{align}
\label{eq:M_G_selection}
    M_\text{G} < 12.7\,.
\end{align}
We also apply a color cut at the blue end to control the uncertainty of photometric mass and age determination
\begin{align}
\label{eq:color_selection}
    G_\text{BP} - G_\text{RP} > -0.6\,.
\end{align}
The selection region on the H--R diagram and 1395 selected white dwarfs\footnote{A catalog of all selected white dwarfs is available on VizieR and on the website: \DWDmergersurl.} are shown in Figure~\ref{fig:HR_merger}. 

We divide our selected sample into five mass bins, based on the aforementioned photometric mass (assuming C/O-core below 1.07~$M_\odot$ and O/Ne-core above it). The edges of bins are 0.8, 0.9, 1.0, 1.1, 1.2, and 1.28~$M_\odot$. If some white dwarfs heavier than 1.07~$M_\odot$ are believed to still hold C/O cores instead of O/Ne cores, such as massive double-WD merger products \citep[e.g.,][]{Dan_2014}, then for those white dwarfs the mass bins correspond to 0.8, 0.9, 1.0, 1.14, 1.24, and 1.32~$M_\odot$. The sample sizes in these mass bins are 408, 431, 323, 176, and 57, respectively. Because of the absolute-magnitude cut and blue color limit, the photometric-age ranges for the five samples are different. We estimate them to be 0.42, 0.82, 0.86, 0.66, and 0.42~Gyr, respectively. 

We estimate the completeness of our sample with the completeness--magnitude relation, $c(G)$, derived in \citet{Sollima_2019} by randomly selecting 180 regions in the sky and comparing {\it Gaia} DR2 and PanSTARRS DR2 catalogs. His estimate is similar to that given by comparing {\it Gaia} DR2 and \textit{Hubble Space Telescope} images around globular clusters \citep{Arenou_2018}. So, we adopt the relation in Figure 1 of \citet{Sollima_2019} and calculate for each mass bin the average completeness, $1/\overline{c^{-1}(G_i)}$, where $G_i$ represents the $G$-band magnitude of a single star. The resulting completeness is 88\%, 80\%, 79\%, 79\%, and 80\%, respectively. In addition, we find that our quality cuts only exclude less than 5\% of the objects, given the distance and H--R diagram cuts in Equations~\ref{eq:d_selection}--\ref{eq:color_selection}.

Finally, we derive the kinematics of white dwarfs, which are related to the true ages of white dwarfs through the age--velocity-dispersion relation. Because {\it Gaia} does not provide radial velocity information for white dwarfs due to the narrow wavelength coverage of its spectrometer \citep{GaiaCollaboration_2016}, we focus on the two components of transverse velocity $\bvT=(\vL,\vB)$:
\begin{align}
    \vL &= \frac{\mu_l-(A\cos 2l + B )\cos b}{\varpi}\,,\\
    \vB &= \frac{\mu_b+A\sin 2l \sin b \cos b}{\varpi}\,,
\end{align}
where $\mu_l$ and $\mu_b$ are the proper motion in the Galactic longitude and latitude directions, and $A$ and $B$ are the Oort constants taken from \citet{Bovy_2017}.

\begin{figure}
    \centering
    \includegraphics[width=\columnwidth]{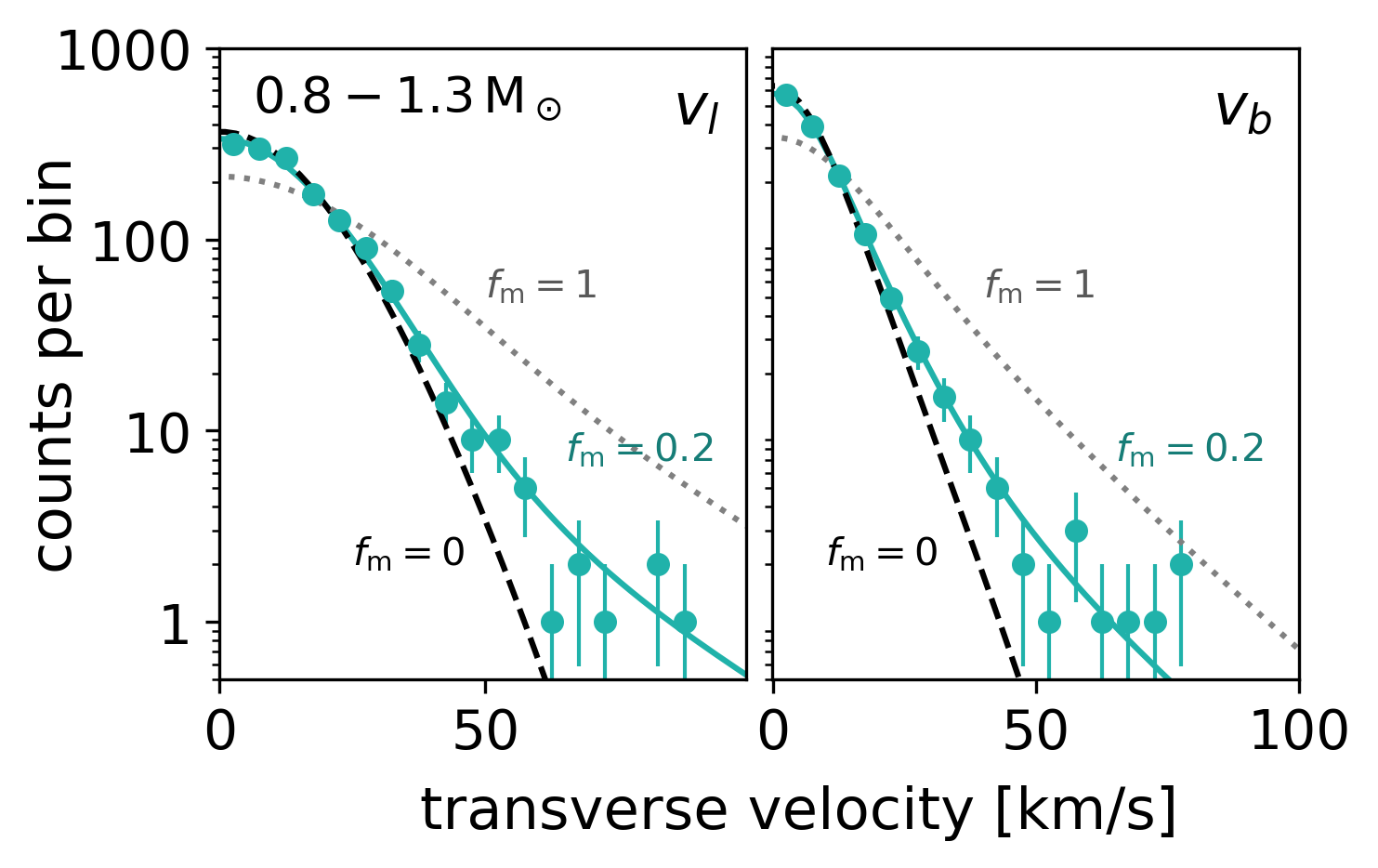}
    \caption{\label{fig:GOF} Velocity distribution of our white dwarf sample. We show the sample of white dwarfs from all five mass bins (0.8--1.3~$M_\odot$) as an example. $\vL$ and $\vB$ in the left and right panel of the figure means the Galactic longitude and latitude components of the transverse velocity. We present the observed histograms of the absolute values of $\vL$ and $\vB$ in 20 bins between 0 and 100~km s$^{-1}$ and Poisson errors. We also show the theoretical velocity distributions for $f_\text{m}$ = 0, 1, and the average of best-fitting values  weighted by the sample size in each mass bin, which is about 0.2. Note that the $y$-axes are in logarithmic scale.}
\end{figure}

\section{Model}
\label{sec:method}

Our goal is to measure the amount of double-WD merger products among high-mass white dwarfs using the kinematic information. According to the age--velocity-dispersion relation, a group of stars with older true stellar age ($\tau$) has higher velocity dispersion. On the other hand, one can derive the photometric isochrone age ($\tau_\text{phot}$) of white dwarfs from the H--R diagram. If a white dwarf evolves in isolation, $\tau_\text{phot}$ should be equal to $\tau$, whereas if it originates from a double-WD merger event, then an age discrepancy,
\begin{equation}
\label{eq:tau}
    \Delta t \equiv \tau - \tau_\text{phot}\,,
\end{equation}
will be created from binary evolution. In general, a white dwarf produced from binary evolution may have positive or negative $\Delta t$, but for double-WD mergers with high total mass, the discrepancy is almost always positive \citep{Temmink_2019}.  So, for a given $\tau_\text{phot}$, double-WD merger products are older and have higher velocity dispersion.

Following \citet{Wegg_2012} and \citet{Cheng_2019}, we assume that the double-WD merger `resets' the white dwarf back to a sufficiently high temperature, so that the real cooling time is equal to the photometric cooling time. Then, $\Delta t$ can also be expressed as the difference of pre-cooling times between the two evolutionary scenarios, $\Delta t = (\tau-t_\text{cool}) - (\tau_\text{phot}-t_\text{cool})$, where $t_\text{cool}$ is the cooling time, and the first item $\tau-t_\text{cool}$ is sometimes called the `delay time' of double-WD merger. It has been widely used that the distribution of the age discrepancy, $p(\Delta t)$, for double-WD mergers with high total masses is approximately a power law, i.e., $p(\Delta t)\approx \Delta t^{-1}$ \citep{Maoz_2010}, because the binary delay time $\tau-t_\text{cool}$ is dominated by the double-WD phase when the orbit shrinks due to gravitational-wave emission, and the single-star pre-cooling time $\tau_\text{phot}-t_\text{cool}$ is negligible. However, in our mass ranges, none of the two statements are valid. So, we use more realistic distributions for $\Delta t$, with the binary delay-time distribution, $p(\tau-t_\text{cool})$, obtained from binary population synthesis (see \ref{app:DTD} for details) and the values of $\tau_\text{phot}-t_\text{cool}$ from Section~\ref{sec:data}.

We consider our white dwarf sample as a mixture of two populations: singly-evolved white dwarfs and double-WD merger products\footnote{The high-mass white dwarfs originating from other types of mergers such as main-sequence and giant star mergers have much shorter age discrepancy $\Delta t$ than that of double-WD mergers. So, in terms of kinematics, we treat the merger products of other types the same as singly-evolved white dwarfs.}, with fractions $1-f_\text{m}$ and $f_\text{m}$, respectively. If $f_\text{m}$ is higher, the tail of the velocity distribution will also be higher, because the double-WD merger products are on average older. For the velocity distribution, we assume that stars with the same true age $\tau$ have a Gaussian velocity distribution $\mathbfit{v}\sim \mathcal{N}(\mathbfit{v_0}(\tau),\mathbf{\Sigma}(\tau))$ relative to the Sun \citep[e.g.,][]{Binney_2008}. The size of this Gaussian distribution is determined by $\tau$ through the age--velocity-dispersion relation, and the center of this Gaussian distribution is determined by the solar motion and the age-dependent asymmetric drift ($\sigma_{U}(\tau)^2/80\,\text{km s}^{-1}$). 

We use the same Bayesian framework as constructed by \citet{Cheng_2019} to infer the fraction $f_\text{m}$ from the photometric age $\tau_\text{phot}$ and transverse velocity $\bvT$ of each white dwarf. This model uses the conditional probability of $\bvT$ given $\tau_\text{phot}$ and other observables as the likelihood function, and thus it eliminates spatial selection biases. In the model, we set $f_\text{m}$ and the solar motion as free parameters and adopt the best-fitting age--velocity-dispersion relation in \citet{Cheng_2019}, a flat star-formation history in our sample volume, and the delay-time distribution of double-WD mergers shown in the \ref{app:DTD}. We do not need to model the `extra cooling delay' included in \citet{Cheng_2019} because this delay has no effect in our selected region. We do not include white dwarf kick effects in our model, because for single-evolved white dwarfs, the kick velocity during the white dwarf formation is less than 1~km s$^{-1}$ \citep{El-Badry_2018}, and for double-WD mergers, the kick velocity during merger is a few km s$^{-1}$ \citep[e.g.,][]{Dan_2014}, which have only a tiny contribution to the increase of velocity dispersion compared to the contribution from the binary-evolution delay.

\begin{figure}
    \centering
    \includegraphics[width=\columnwidth]{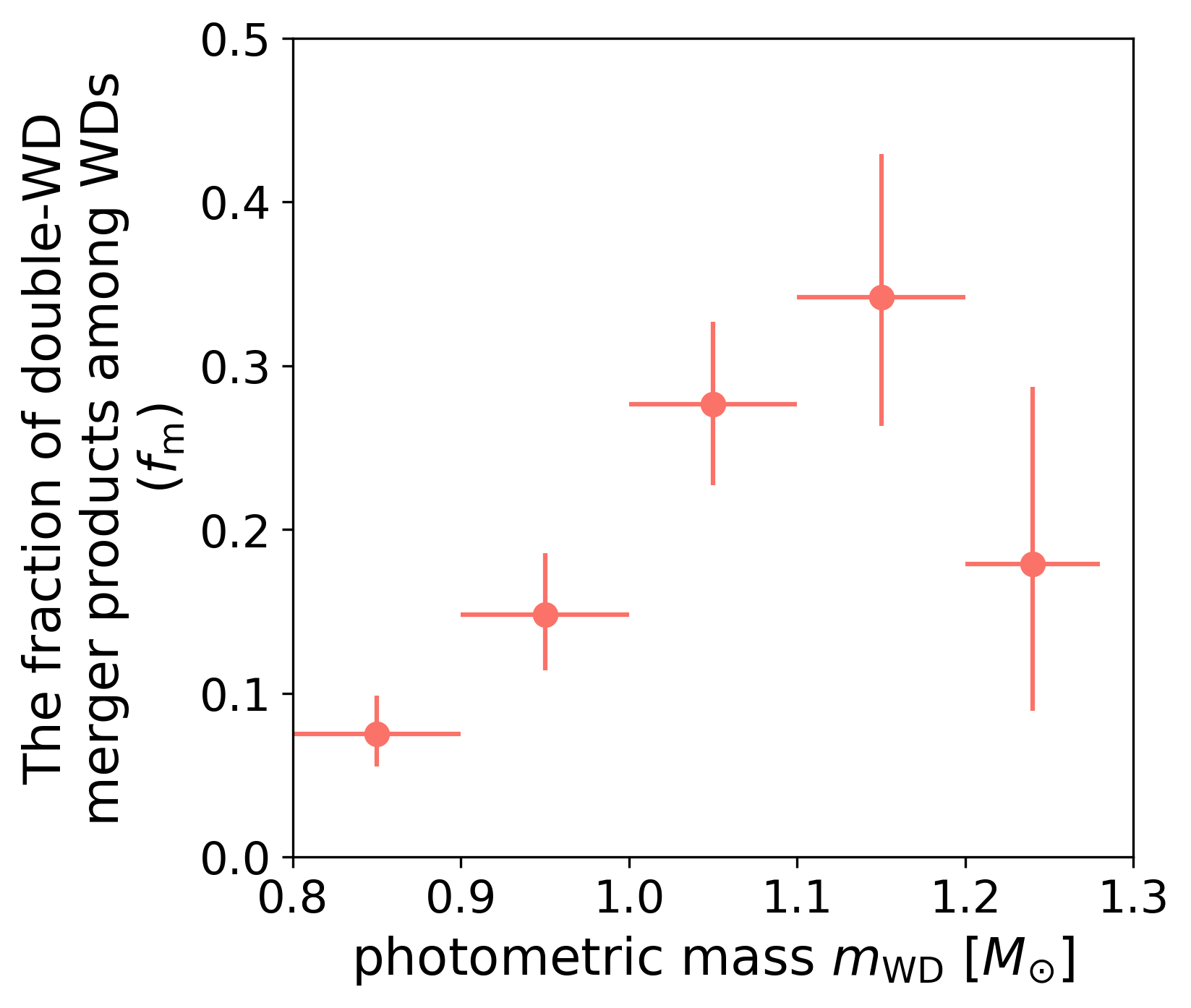}
    \caption{\label{fig:f_merger} Our estimates for the fraction of double-WD merger products among high-mass white dwarfs, in five bins of white dwarf photometric mass. The sample sizes in these mass bins, from lower to higher masses, are 408, 431, 323, 176, and 57, respectively.}
\end{figure}

\begin{figure*}
    \centering
    \includegraphics[width=0.85\textwidth]{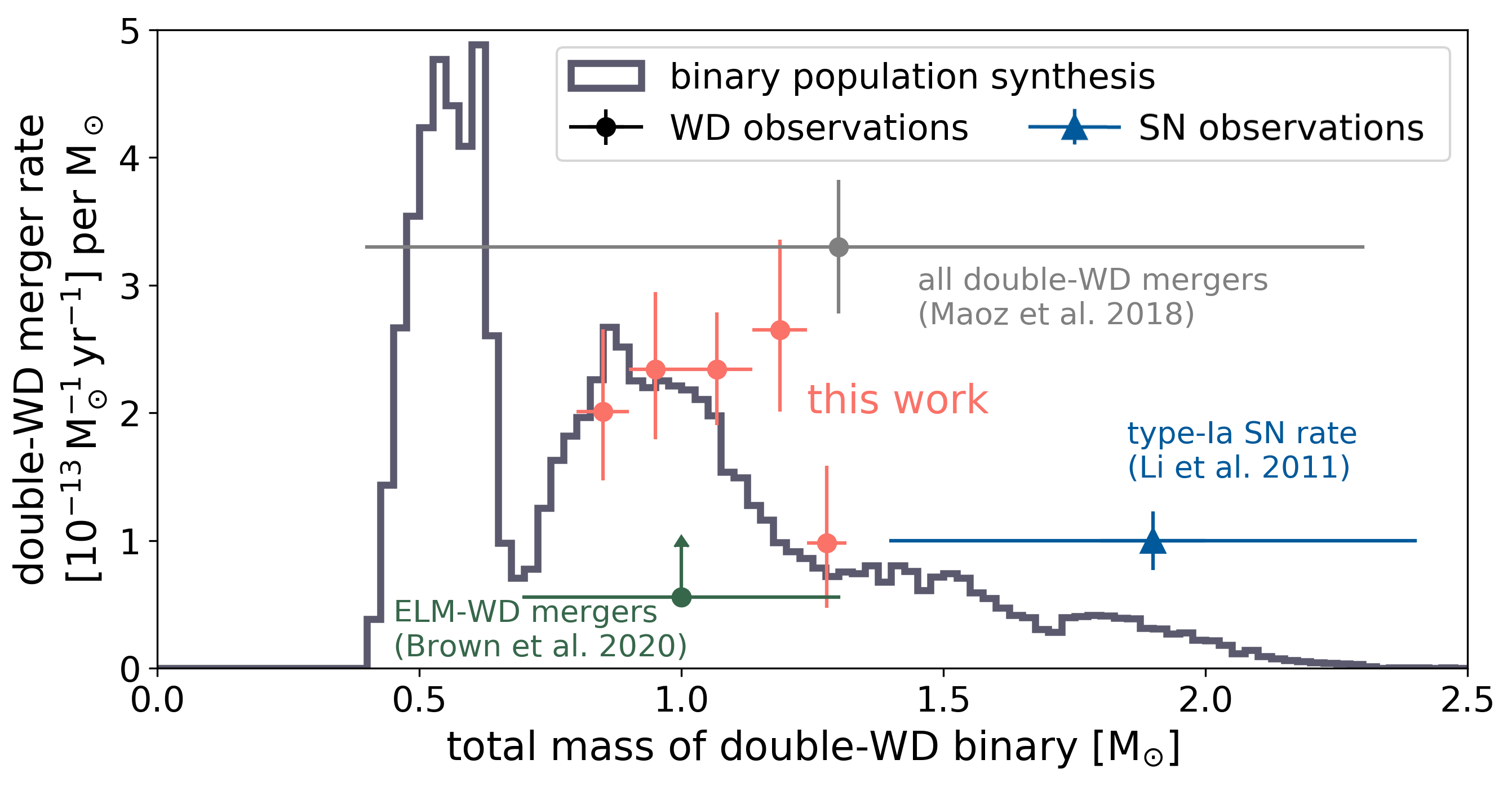}
    \caption{\label{fig:merger_rate} A comparison of the observed and simulated double-WD merger rate. The red data points with error bars are our observational estimates based on double-WD merger products. The histogram shows binary population synthesis results. Other data points show estimates in the literature based on the orbital distribution of observed double-WD systems: the light-grey one is an estimate for all double-WD mergers \citep{Maoz_2018}, and the green one is for systems with at least one extremely low-mass (ELM) WD \citep{Brown_2020}, which provides a lower limit of the merger rate. The blue data point shows the observed SN Ia rate. Comparisons between a data point and the histogram should be made in terms of the area under the horizontal `error bar' of the data point and the area under the histogram in the same mass range.}
\end{figure*}

\begin{deluxetable*}{c|ccc}[t]
    \tablecaption{\label{tab:merger_rate}Measurements of the Double-WD (DWD) Merger Rate and Type Ia Supernovae (SNe Ia) Rate}
    \tablehead{
        \colhead{Reference}&\colhead{Event Rate}&\colhead{Event Type}&\colhead{Based on}\\
        \colhead{}&\colhead{(10$^{-13}\,M_\odot^{-1}\,\text{yr}^{-1}$)}&\colhead{}&\colhead{}
    }
    \startdata
    This work & 0.20 $\pm$ 0.06&DWD mergers that produce 0.8--0.9 $M_\odot$ WDs & Merger products among single WDs\\
    &0.23 $\pm$ 0.06&DWD mergers that produce 0.9--1.0 $M_\odot$ WDs\\
    &0.32 $\pm$ 0.06&DWD mergers that produce 1.0--1.14 $M_\odot$ WDs\\
    &0.28 $\pm$ 0.07&DWD mergers that produce 1.14--1.24 $M_\odot$ WDs\\
    &0.07 $\pm$ 0.04&DWD mergers that produce 1.24--1.32 $M_\odot$ WDs\\
    \hline
    This work (summed)&1.1 $\pm$ 0.3&DWD mergers that produce 0.8--1.32 $M_\odot$ WDs & Merger products among single WDs\\
    \citet{Maoz_2018} & 6.3 $\pm$ 1.0 & All DWD mergers & Orbital distribution of DWD systems\\
    \citet{Brown_2020} & 0.3 $\pm$ 0.2 & DWD mergers with at least one ELM WD& Orbital distribution of DWD systems\\
    \hline
    \citet{Li_2011} & 1.0 $\pm$ 0.3 & SNe Ia in a Milky-Way-like galaxy & Extra-galactic SNe
    \enddata
\end{deluxetable*}

\section{Results and Discussions}
\label{sec:results}

\subsection{Constraints on the fraction of merger products}
\label{sec:f_merger}

With a thirty-time larger sample of high-mass white dwarfs selected from {\it Gaia} DR2, we are allowed to go beyond \citet{Wegg_2012} and set strong constraints on $f_\text{m}$. Figure~\ref{fig:GOF} illustrates the transverse-velocity distribution of white dwarfs in our sample. The clear velocity excess is strong evidence for the existence of double-WD merger products. For clarity we only show the distribution of the whole sample, i.e., the combination of all five mass bins, but similar results can be found in each single mass bin, too. In Figure~\ref{fig:f_merger}, we show our estimate of $f_\text{m}$ in each mass bin. We find that the fraction of double-WD mergers in our mass range of 0.8--1.3~$M_\odot$ varies from 10\% to 35\%, with an average of about 20\%. This fraction is roughly constant as a function of mass, though declines at the two end are suggested.

To test the robustness of our results, we check for the influence of sample selection, the adopted star-formation history, and the adopted age--velocity-dispersion relation in our model. We found that a different distance cut such as 200 or 300~pc cuts and a linearly decreasing star-formation history with 5 times higher rate at 11~Gyrs ago have less than 20\% \emph{fractional} influence on the estimate of $f_\text{m}$. For the influence of the age--velocity-dispersion relation, our results are mostly influenced by the 0--4~Gyr part, where the delay-time distribution is peaked. Adopting the high velocity dispersion from \citet{Just_2010} as used by \citet{Wegg_2012} will reduce $f_\text{m}$ by a factor of 2, but given the observational constraints from both main-sequence stars \citep[e.g.,][]{Nordstrom_2004} and white dwarfs \citep{Cheng_2019}, such high values of velocity dispersion are unlikely. The effect of adopted delay-time distribution can be seen from the comparison between our results in the mass range of 1.08--1.23~$M_\odot$ and that of \citet{Cheng_2019}: adopting a power-law delay-time distribution leads to a result about 30\% lower. So, we estimate the \emph{fractional} systematic error of our results as 30\% (a factor of 0.7--1.3). Our estimate of $f_\text{m}$ is consistent with population synthesis results \citep{Temmink_2019}.

\subsection{Double-WD merger rate}
\label{sec:merger rate}

The fraction of double-WD merger products ($f_\text{m}$) obtained in Section~\ref{sec:f_merger} can be translated into double-WD merger rates. Because our sample is nearly volume limited, the merger rate in each mass bin can be estimated by
\begin{equation}
    \text{merger rate}=\frac{f_\text{m}\cdot N}{m_{\star} \cdot \Delta \tau_\text{phot} \cdot c}\,,
\end{equation}
where $N$ is the sample size of each mass bin (listed in the caption of Figure~\ref{fig:f_merger}), $m_\star$ the stellar mass of the Milky Way within 250~pc. $\Delta \tau_\text{phot}$ and $c$ are the photometric-age range and sample completeness of each mass bin, which are estimated in Section~\ref{sec:data}. The stellar mass $m_\star$ is estimated to be $4.1\times10^6\,M_\odot$, using the local stellar mass density $\rho_\star=0.083\,M_\odot\,\text{pc}^{-1}$ \citep{McMillan_2011} and a scale-height 300~pc of the disc. 
We list our estimate of the current double-WD merger rate in each mass bin in Table~\ref{tab:merger_rate}. The total merger rate in our mass range (0.8--1.3 $M_\odot$) amounts to $1.1\times 10^{-13} \,M_\odot^{-1}\,\text{yr}^{-1}$. To make comparison with other measurements easier, we also divide these values by their corresponding mass ranges and show the results in Figure~\ref{fig:merger_rate}. Note that the mass range of each bin here is slightly different from that in Figure~\ref{fig:f_merger}, because we adopt the photometric masses derived from CO white dwarf models (see Section~\ref{sec:data} and Table~\ref{tab:merger_rate} for details).

There are several factors that may lead us to underestimating the merger rate within our mass range. For example, we take the mass of the merger products as the total mass of the original double-WD binary. This is true for CO-CO white dwarf mergers \citep[e.g.,][]{Dan_2014} but not for He-CO white dwarf mergers, which may lose significant amount of mass during the R Coronae Borealis phase and produce white dwarfs with 0.6--0.7~$M_\odot$ \citep{Schwab_2019}. So, we are likely to underestimate the merger rate of systems with original mass below 1.0~$M_\odot$, where He-CO white dwarf mergers become important. Similarly, we will miss explosive and disruptive merger events if there be any such events in our mass range, and events that result in extremely magnetic and faint white dwarfs \citep{Bhattacharya_2018} if there be such objects produced.

In Figure~\ref{fig:merger_rate} we also show the merger rate from binary population synthesis, with a flat star-formation history assumed. If a decreasing star-formation rate was assumed, as the delay-time distribution is low for mergers with a long delay time, the synthesis would predict a lower current merger rate than plotted in Figure~\ref{fig:merger_rate}. Details of the population synthesis are shown in \ref{app:DTD}. We find that the synthesized merger rates are close to our observational estimates without any tuning of parameters.
Note that in our analysis of {\it Gaia} white dwarfs, we only use the \emph{distribution} of the delay time but never use the total merger rate information from the population synthesis. So, the match between the observed and synthesized merger rate is not a circular argument but rather a validation of our understanding of binary evolution.

Then, we compare our results with other estimates of the double-WD merger rate in the literature. While we count the products of mergers, other estimates are obtained by observing pre-merger systems and predicting the merger rate.
\citet{Maoz_2012, Maoz_2018}, \citet{Badenes_2012}, and \citet{Maoz_2017b} extrapolate the orbital distribution of double-WD binaries to estimate the total double-WD merger rate, with an up-to-date estimate being (6.3~$\pm$~1.0) $\times$ 10$^{-13}\,M_\odot^{-1}\,\text{yr}^{-1}$. \citet{Brown_2016b, Brown_2020} estimate the merger rate of double-WD binaries with at least one extremely low-mass (ELM; $<$0.3~$M_\odot$) white dwarf to be 2~$\times$~10$^{-3}\,\text{yr}^{-1}$ in the milky way, corresponding to 0.3~$\times$~10$^{-13}\,M_\odot^{-1}\,\text{yr}^{-1}$, with a 110\% uncertainty including 70\% statistical uncertainty. In Table~\ref{tab:merger_rate} we list these values. In Figure~\ref{fig:merger_rate} we assign reasonable mass ranges to these measurements and present the results. For the result from \citet{Maoz_2018}, we assign 0.4--2.3~$M_\odot$ according to the mass distribution in our binary population synthesis, and for the result from \citet{Brown_2020}, we assign 0.7--1.3~$M_\odot$ according to the mass distribution of ELM binaries \citep{Brown_2016a}. All data points on Figure~\ref{fig:merger_rate} should be understood as the averaged merger rate within the assigned mass range. As these measurements address the merger rates of systems in different mass ranges, one cannot compare them directly. But, if we are allowed to use the mass distribution from binary population synthesis to scale these estimates, we will find that the merger rate obtained by \citet{Maoz_2018} is about 2--3 times of our estimates, and the estimate from \citet{Brown_2020} is consistent with our results, as illustrated in Figure~\ref{fig:merger_rate}. As discussed in \citet{Maoz_2018}, if the merger rates are as high as their estimate, almost all high-mass white dwarfs will need to be double-WD merger products, which is hard to believe given the velocity distribution we observe. Nevertheless, it is noticeable that the observational constraints of the double-WD merger rate from different methods have converged to within a factor of a few.

In summary, our estimates of the double-WD merger rate add significant precision and mass resolution to our knowledge of the double-WD merger rate and provide a validation for current binary population synthesis.

\subsection{Implication for type Ia supernovae}

Type Ia supernovae are important distance indicators, element factories, interstellar medium heaters, and cosmic-ray accelerators, but their progenitors remain unclear \citep[e.g.,][]{Maoz_2014}.
The double-WD merger is a promising scenario of type Ia supernova \citep[e.g.,][]{Iben_1984, Webbink_1984, Tutukov_1992, Maoz_2010, Mennekens_2010, Sato_2015, Liu_2017, Shen_2018a, Shen_2018b, Perets_2019}. The comparison between double-WD merger rate and the type Ia supernova rate is a critical test from this scenario. When a flat star-formation history is assumed, our population synthesis (\ref{app:DTD}) provides a merger rate of about 0.3~$\times$~10$^{-13}\,M_\odot^{-1}\,\text{yr}^{-1}$ for super-Chandrasekhar double-WD systems, which is about 1/7 of the total synthesized double-WD merger rate and consistent with previous studies \citep[e.g.,][]{Ruiter_2009, Yungelson_2017}. For the D$^6$ (dynamically driven double-degenerate double-detonation) scenario \citep[e.g.,][]{Shen_2018a}, a lower rate is obtained, because it requires in general higher total mass of the system \citep[see figure 2 of][]{Shen_2017}.

On the other hand, the observed type Ia supernova rate for a Milky-Way-like galaxy (Sb-Sbc type) is (1.0~$\pm$~0.3) $\times$ 10$^{-13}\,M_\odot^{-1}\,\text{yr}^{-1}$ \citep{Li_2011}, or 1.3~$\times$~10$^{-3}\,M_\odot^{-1}$ in terms of a time-integrated rate \citep{Maoz_2017a}. This is close to, though 2--3 times higher than, the synthesized rate for the super-Chandrasekhar and D$^6$ double-WD merger scenario. As discussed in Section~\ref{sec:merger rate}, our estimates of the double-WD merger rate within 0.8--1.3~$M_\odot$ are in agreement with population synthesis results. So, if we are allowed to extrapolate the merger rate to higher masses according to the mass distribution of mergers from simulations, then
\begin{itemize}
    \item our measurements support the idea that double-WD mergers can contribute a significant fraction to type Ia supernovae; 
    \item if all type Ia supernovae come from double-WD mergers, it seems that there exist other explosion mechanisms whose requirement on the total mass of the binary is lower than that of the Chandrasekhar- and D$^6$ explosion models.
\end{itemize}

\section{Conclusion}
\label{sec:conclusion}

The merger of two white dwarfs in a close binary system may result in a new white dwarf with higher mass. Therefore, among the high-mass white dwarfs observed today, a fraction should come from double-WD mergers. Experiencing binary evolution, these merger products have older true ages than their photometric isochrone ages. According to the age--velocity-dispersion relation in the Milky-Way disc, older stars have higher velocity dispersion. So, the fraction of double-WD merger products ($f_\text{m}$) can be estimated from the velocity distribution of high-mass white dwarfs. 

We select a homogeneous sample of high-mass white dwarfs (0.8--1.3~$M_\odot$, $d<$~250 pc) from {\it Gaia} DR2, which includes 1395 objects. Our sample is about thirty times larger than that of a previous study with a similar idea \citep{Wegg_2012}.
We infer $f_\text{m}$ in five mass bins using a Bayesian model of white dwarf transverse velocities. We find
\begin{enumerate}
    \item about 20\% of 0.8--1.3 $M_\odot$ white dwarfs originate from double-WD mergers;
    \item the corresponding double-WD merger rates in our mass range add up to 1.1~$\times$~10$^{-13}\,M_\odot^{-1}\,\text{yr}^{-1}$.
\end{enumerate}
We show $f_\text{m}$ and the merger rate as a function of mass in Figures~\ref{fig:f_merger} and \ref{fig:merger_rate}, respectively. We estimate our systematic error to be within 30\%, i.e., a factor of 0.7--1.3. Our results are in good agreement with the predictions from binary population synthesis (see \ref{app:DTD} for setting details).

Our estimates add significant precision and mass resolution to our knowledge of the double-WD merger rate. If it is allowed to extrapolate the estimates to a higher mass range, our results suggest that double-WD mergers can contribute to a significant fraction of type Ia supernovae.

In a few years, the increasing astrometric and photometric precision provided by future {\it Gaia} data releases and the radial velocity measurements of white dwarfs by future surveys such as SDSS-V \citep[e.g.,][]{Kollmeier_2017} will enlarge the available sample size of high-mass white dwarfs and allow for even tighter constraints. We are starting to be able to reliably and precisely compare the observed double-WD merger rates with binary population synthesis, which will shed light upon the progenitor problem of type Ia supernovae.

\section*{Acknowledgments}

We thank the referee Dan Maoz for providing many valuable suggestions to improve the quality of our draft. 
S.C. thanks Siyu Yao for her constant encouragement and inspiration.
J.C. would like to acknowledge his support from the National Science Foundation (NSF) through grant AST-1614933.
B.M. thanks the David and Lucile Packard Foundation.
S.T. acknowledges support from the Netherlands Research Council NWO (grant VENI [nr. 639.041.645]).

This work has made use of data from the European Space Agency (ESA) mission {\it Gaia} (\url{https://www.cosmos.esa.int/gaia}), processed by the {\it Gaia} Data Processing and Analysis Consortium (DPAC, \url{https://www.cosmos.esa.int/web/gaia/dpac/consortium}). Funding for the DPAC has been provided by national institutions, in particular the institutions participating in the {\it Gaia} Multilateral Agreement.

\software{astropy package \citep{AstropyCollaboration_2013, AstropyCollaboration_2018}, emcee \citep{Foreman-Mackey_2013}, numpy \citep{oliphant2006guide},  matplotlib \citep{Hunter_2007}, SciPy \citep{Virtanen_2019}}

\begin{appendix}
\section{The binary population synthesis}
\label{app:DTD}

\begin{figure}
    \centering
    \includegraphics[width=\columnwidth]{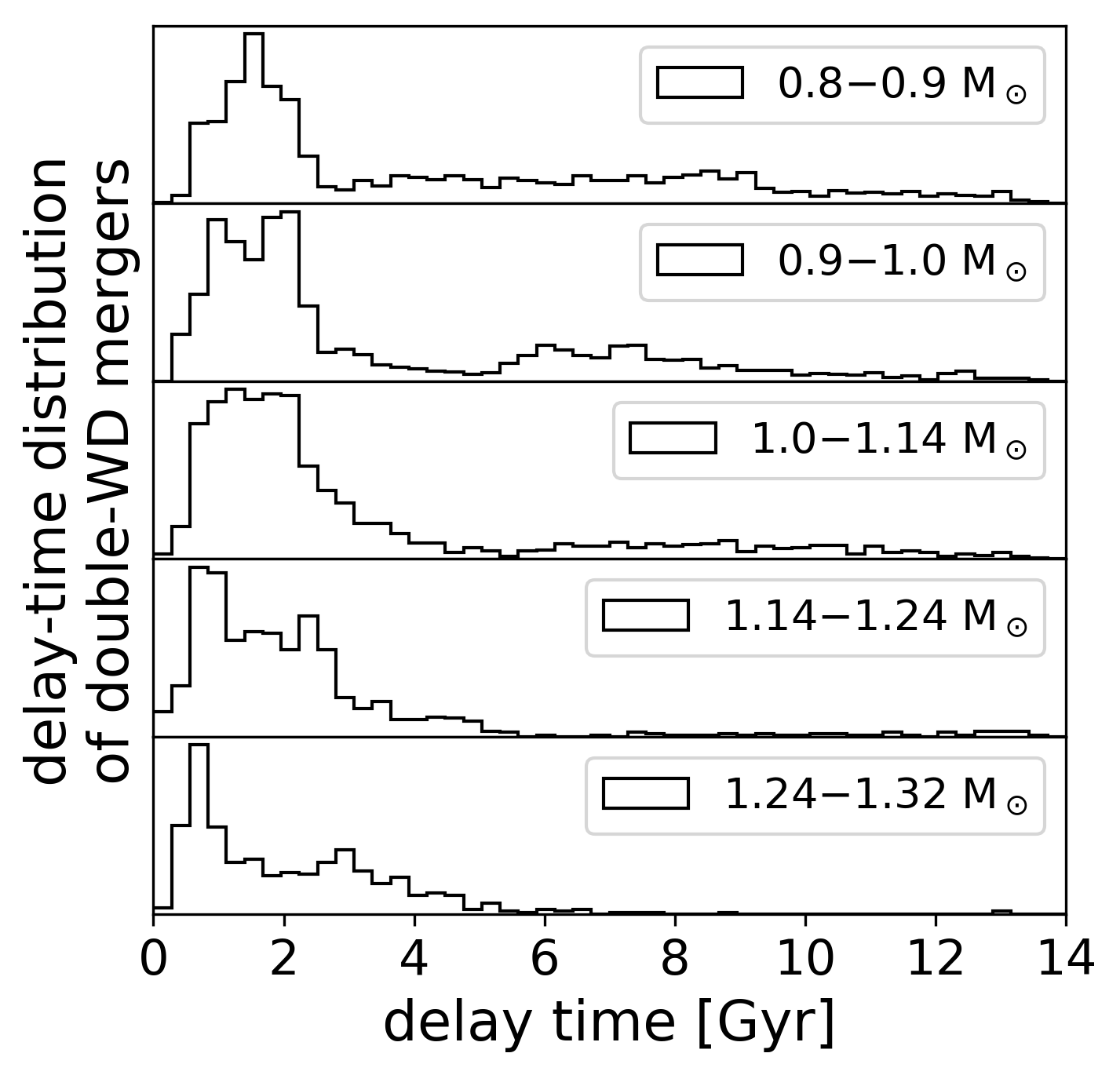}
    \caption{\label{fig:DTD} Delay-time distributions of double-WD mergers used in our model. These distributions are generated from binary population synthesis. The $x$-axis is the delay time of binary evolution, i.e., $\tau-t_\text{cool}$ for the resulting white dwarf. The y-axis is in linear scale and normalized to their maximum values. We input to our model the shapes of these five distributions as probability distribution and do not use the information from their normalization.}
\end{figure}

Here we describe the binary population synthesis that we use in this paper to derive the delay-time distribution. 
The models are synthesized using the binary population synthesis code \texttt{SeBa} \citep{Por96, Toonen_2012}. The models are identical to the default models used in \cite{Toonen_2017}. We have adopted a Kroupa initial mass function \citep{Kro93} and a uniform mass ratio distribution between 0 and 1 \citep{Raghavan_2010, Duchene_2013}. Furthermore, we assume a uniform distribution in the logarithmic semi-major axis up to $10^6\,R_{\odot}$ \citep{Abt83}, and a thermal distribution of eccentricities between 0 and 1 \citep{Heg75}. 

One of the main sources of uncertainty in the synthetic populations \citep{Toonen_2014} is a phase of unstable mass transfer, i.e., the common-envelope (CE) phase \citep[for a review, see][]{Iva13}. Similar to \cite{Toonen_2017}, we apply the `$\gamma\alpha$' model. This model reproduces the mass ratio distribution \citep{Toonen_2012} and number density \cite{Toonen_2017} of double-WD systems best. In the `$\gamma\alpha$' model, we apply the classical ($\alpha$-)CE that is based on energy conservation \citep{Webbink_1984}, and the ($\gamma$-)modelling that is based on a balance of angular momentum \citep{Nel00}. Regarding the former the parameters $\alpha\lambda=2$ describe how efficient orbital energy can be used to unbind the envelope and how strong the envelope is bound to the donor star, and regarding the latter the parameter $\gamma=1.75$ describes the efficiency of angular momentum usage. The $\gamma$-modelling is applied unless the binary contains a compact object or the CE is triggered by a tidal instability.  
We note that for our purpose to compare the merger rate, the delay-time distribution of the `$\gamma\alpha$' model does not significantly differ from that of the model that exclusively adopts the $\alpha$-CE with $\alpha\lambda=2$ \citep[see][]{Toonen_2012}. 

Figure~\ref{fig:DTD} shows the delay-time distributions in five mass bins, which are used in our kinematic analysis. For the synthesized merger rates shown in Figure~\ref{fig:merger_rate}, we in addition assume a 50\% binary fraction of all stars \citep[see also][]{Duchene_2013, Moe17}.

\end{appendix}

\bibliography{double_WD_merger}
\end{CJK*}
\end{document}